\documentclass[runningheads,a4paper]{llncs}

\usepackage{amssymb}
\usepackage{graphicx}

\usepackage{url}
\urldef{\mailsa}\path|{alfred.hofmann, ursula.barth, ingrid.haas, frank.holzwarth,|
\urldef{\mailsb}\path|anna.kramer, leonie.kunz, christine.reiss, nicole.sator,|
\urldef{\mailsc}\path|erika.siebert-cole, peter.strasser, lncs}@springer.com|
\newcommand{\keywords}[1]{\par\addvspace\baselineskip
\noindent\keywordname\enspace\ignorespaces#1}

\newcommand{\vp}{\varphi}
\RequirePackage{color}

\def \figr #1#2#3
{   \begin{figure}[ht]
        \centering\includegraphics[width=#2 \textwidth]{#1.eps}
        {\caption {#3}\label{#1}}
    \end{figure}
}   
\def \myfigures #1#2#3#4#5#6#7#8
{\begin{figure}[h!t]
    \centering\includegraphics[width=#2 \textwidth]{#1.eps}
        \hfill
    \centering\includegraphics[width=#6 \textwidth]{#5.eps}
        \parbox[t]{#4\textwidth}{\caption {#3}\label{#1}}
        \hfill
        \parbox[t]{#8\textwidth}{\caption {#7}\label{#5}}
\end{figure} }

\begin{document}
\mainmatter  

\title{Numerical study of fluxon solutions of sine-Gordon equation under the influence of the boundary conditions}
\author{P.~Kh.~Atanasova\inst{1},
E.~V.~Zemlyanaya\inst{2},
Yu.~M.~Shukrinov\inst{2}}
\authorrunning{Atanasova P.~Kh.,  Zemlyanaya E.~V., Shukrinov Yu.~M.}
\titlerunning{Numerical study of fluxon solutions of sine-Gordon equation \ldots}
\institute{Plovdiv University ``Paisii Hilendarski'', 4000 Plovdiv, Bulgaria
\and
Joint Institute for Nuclear Research,
 141980  Dubna,  Russia }
\maketitle

\begin{abstract}

\emph{The fluxon solutions of a boundary problem for the sine-Gordon equation (SGE) are investigated numerically in dependence on the boundary conditions. Interconnection between fluxon and constant solutions is
analyzed. Numerical results are discussed in context of the long Josephson junction model.
}

\end{abstract}

\section{Introduction}

The sine-Gordon equation is a nonlinear hyperbolic partial differential equation involving
the d'Alembert operator and the sine of the unknown function. The equation reads
\begin{equation}\label{sg_partial_dif}
    u_{tt} - u_{xx} + \sin u = 0\,,
\end{equation}
where $u = u(x, t)$.
It arises in differential geometry and various areas of physics, including applications in relativistic field theory,
Josephson junctions (JJs) or mechanical transmission lines. The stack of coupled JJs describing by system of coupled sine-Gorgon equations is investigating very intensively today \cite{koshelev,hu,krasnov}.

In the framework of the long JJs model, the dynamics of the magnetic flux in a JJ of length $2l$
is described  by the perturbed sine-Gordon equation:
 \begin{equation}\label{dinamic_equation_s}
  \varphi_{xx} - \varphi_{tt} - \alpha \varphi_t = \sin \varphi - \gamma\,,
  \quad t > 0\,, \quad x \in (-l,l)
 \end{equation}
with boundary conditions
\begin{equation}\label{postanovka_gr_usl_s}
  \varphi_x (\pm l,t) = h_e\,,
\end{equation}
where  $\varphi$ is the magnetic flux distribution, $h_e$ -- the external  magnetic field, $\gamma$ -- the external current
and $\alpha \geq 0$ -- the dissipation coefficient.

The aim of this work is a numerical investigation of the properties of the {\it static} fluxon solutions of Eq.(\ref{dinamic_equation_s}) under the influence of the external magnetic field parameter $h_e$ in (\ref{postanovka_gr_usl_s}). Such solutions satisfy the following boundary problem

\begin{equation}\label{sg}
    -\varphi_{xx} + \sin \varphi -\gamma = 0, \quad x \in (-l;l), \quad
    \varphi_x(\pm l) = h_e\,.
\end{equation}

Here, we only consider the case of the JJ length $2l = 10$ with zero external current $\gamma = 0$.
%
\section{Numerical approach}
%
     The static fluxon solutions of Eq.(\ref{dinamic_equation_s}) are obtained numerically, by solving of the boundary problem (\ref{sg}).
     The stability analysis is based on numerical solution of the following Sturm-Liouville problem \cite{galfil_84}
\begin{equation}
\label{slp}
   -\psi_{xx} + q(x)\psi = \lambda \psi, \quad
    \psi_x(\pm l) = 0, \quad q(x) =  \cos \varphi(x,p), \quad p = (l,h_e,\gamma).
\end{equation}
In this approach, the minimal eigenvalue of Eq.(\ref{slp}) $\lambda_0(p) > 0$ corresponds to a stable solution. In case $\lambda_0(p) < 0$ solution $\vp(x,p)$ is unstable. The case $\lambda_0(p) = 0$ indicates a bifurcation with respect to one of parameters $p = (l,h_e,\gamma)$.

The numerical solving of Eq.(\ref{sg}) is based of the continuous analog of Newton's method \cite{pbvzpc07}. At each Newtonian iteration the corresponding linearized problem is solved, on a uniform grids with $1025$ number of nodes, using a three-point
Numerov approximation of the fourth  order accuracy \cite{numerov}.

For numerical solution of the Sturm-Liouville  problem (\ref{slp}) we applied the standard three-point second order finite-difference formulae. First several eigenvalues of the  resulting algebraic three-diagonal eigenvalue problem are obtained by means of the standard EISPACK code. Details of numerical scheme are described in \cite{azbsh10,lncs10,fdm10} for the double sine-Gordon equation.

The known for $h_e = 0$ solutions $M_0$ and $\Phi^1$ are numerically path-followed to non-zero positive $h_e$.
At each $i$th step of the numerical continuation we analyze the stability of solution $\vp(x,h_e^{(i)})$ and calculate  the following physical characteristics:
\begin{itemize}
  \item full magnetic flux of the distribution $\Delta\vp^{(i)}\vp = \vp(l,h_e^{(i)}) - \vp(-l,h_e^{(i)})$;
  \item quantity $N$ denoted  ``number of fluxons'' in \cite{azbsh10} and determined as follows
\begin{equation}
N[\vp(x,h_e^{(i)})] = \frac{1}{2l \pi} \int \limits_{-l}^l \varphi(x,h_e^{(i)})\,dx.
\end{equation}
\end{itemize}
Note, since each solution $\varphi$ of Eq.(\ref{sg}) is defined with an accuracy $2k\pi$ ($k \in \mathbf{Z}$) then the value $N[\varphi]$ is also defined with accuracy $2k$. The arbitrariness at the choice of integer number $k$ can be used for the ``concordance'' of the value $N$
 with the value of the full magnetic flux $\Delta\vp$ according to the condition
\begin{equation}\label{n_deltaphi}
    \left|N[\vp] - \Delta\vp/2\pi\right| \rightarrow \min.
\end{equation}

Below, as in \cite{azbsh10,lncs10,fdm10,nano10}, solutions $\vp$ with $n=N$ where N satisfies Eq.(\ref{n_deltaphi}) are denoted $\vp^n$.

The crossing through the turning points in the numerical continuation (where the direction of the moving along the curve $\Delta\vp(h_e)$ changes as we follow on the new branch) was organized as in \cite{continuation}. The turning points are identified with help of the relation that is tested at each $i$th step of numerical continuation:
\begin{equation}
\left| \frac{ {h_e}^{(i)} - {h_e}^{({i-1})}}
{\Delta\vp^{(i)} - \Delta\vp^{(i-1)}} \right| < \varepsilon.
\label{epsilon}
\end{equation}
where $\varepsilon>0$ is small known quantity.
Note that (\ref{epsilon}) is a simple approximation of equality $dh_e/d\Delta\vp=0$ that is valid at the turning points of the curve   $\Delta \vp(h_e)$. In case we run into a turning point the sign of $h_e$-increment should be changed.
At each step of the numerical continuation, the initial guess for Newtonian process was chosen in the form
\begin{equation}
\vp(h_e^{(i+1)}) = \vp(h_e^{(i)}) + (h_e^{(i+1)} - h_e^{(i)}) \cdot \frac
{\vp(h_e^{(i)}) - \vp(h_e^{(i-1)})}
{h_e^{(i)} - h_e^{(i-1)}}
\label{predict}
\end{equation}
that prevents the continuation from reversing to the previous branch of $\Delta\vp(h_e)$.
%
\section{Numerical results} \label{numerical_results}
%
 \myfigures{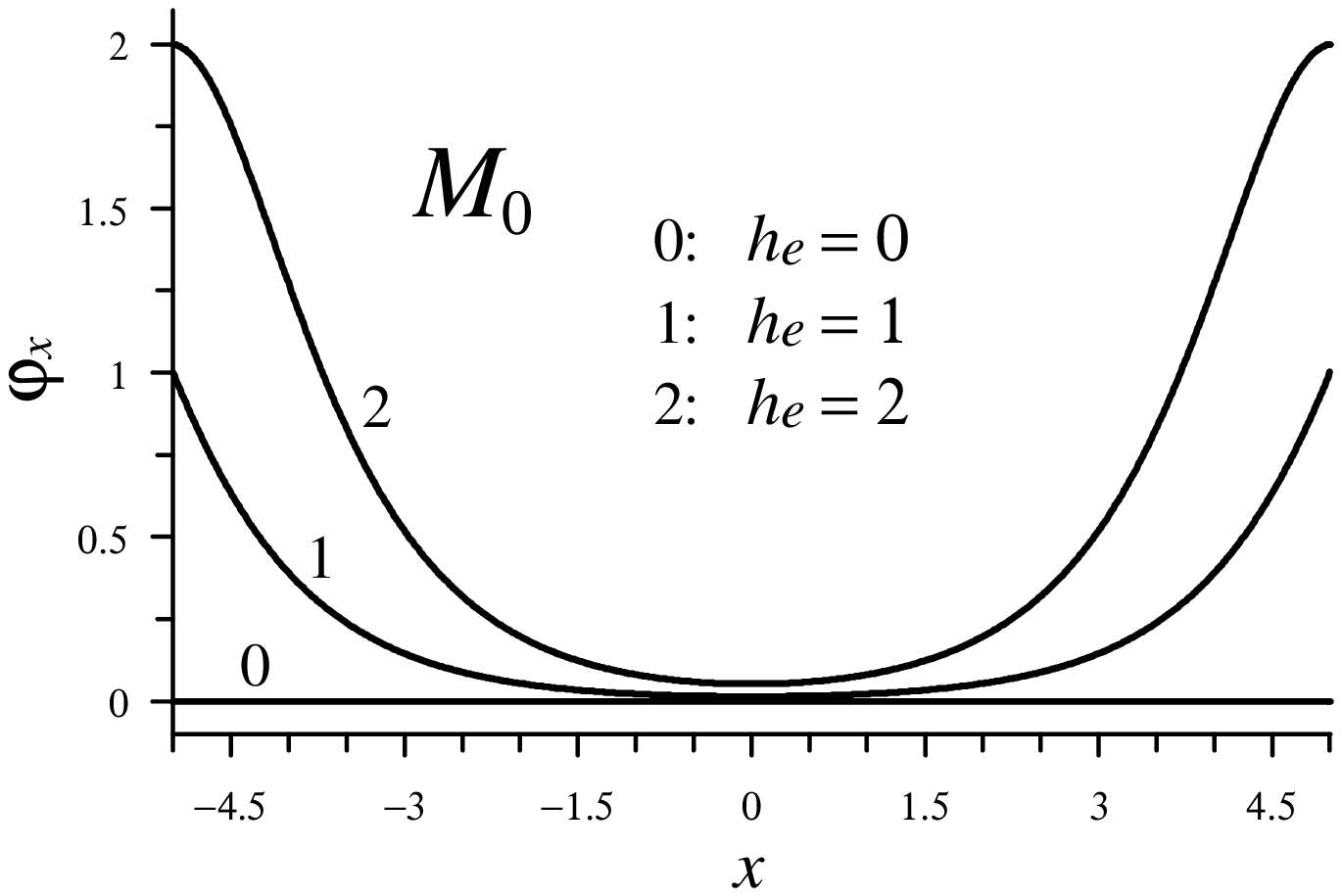}{0.48}{Internal magnetic field distribution $\vp_x(x)$ associated with the state $M_0$ for several values of the magnetic field $h_e$.\label{solsM0}}{0.48}
{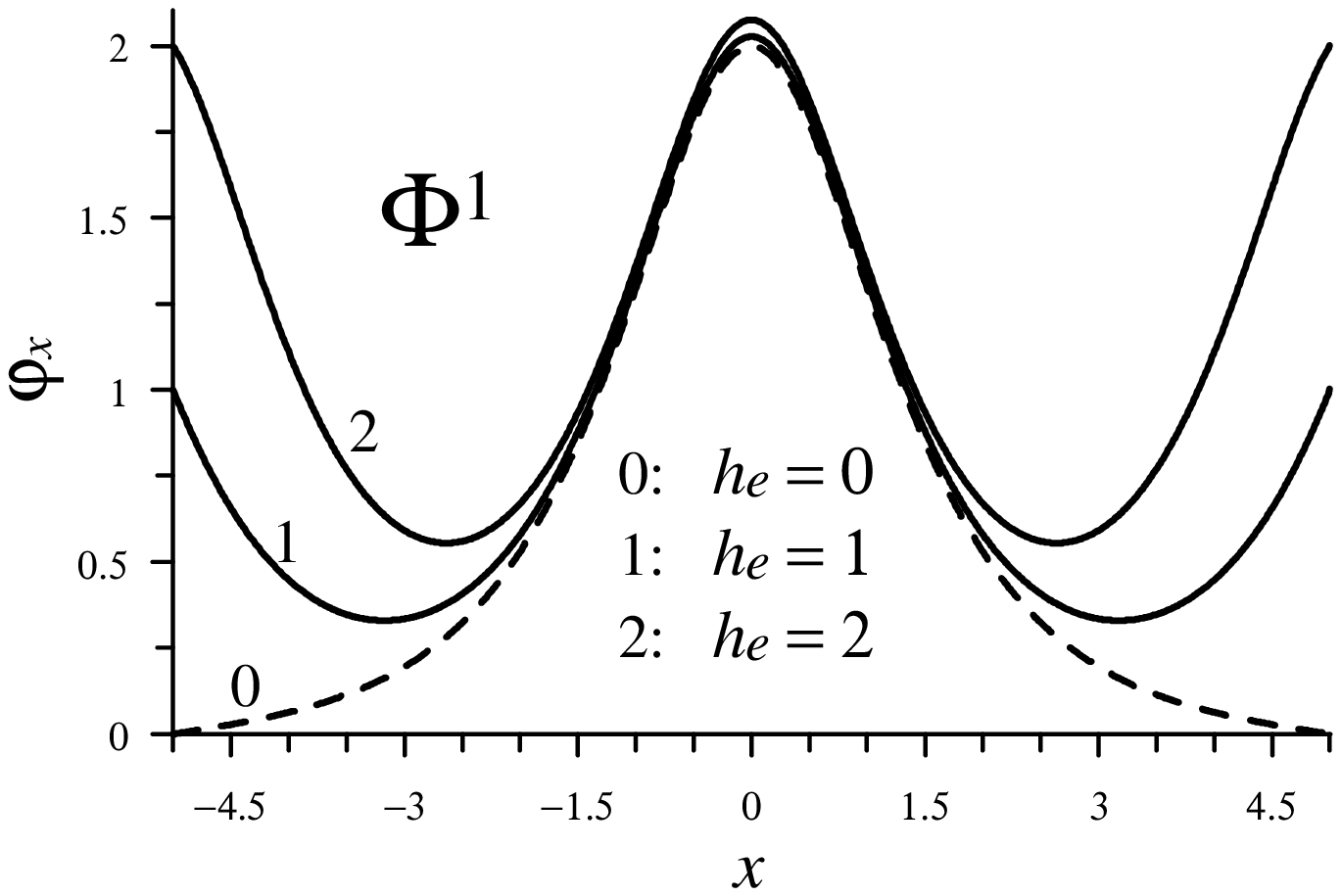}{0.48}{Internal magnetic field distribution $\vp_x(x)$ associated with the state $\Phi^1$ for several values of magnetic field $h_e$.\label{solsF1}}{0.48}

%
\figr{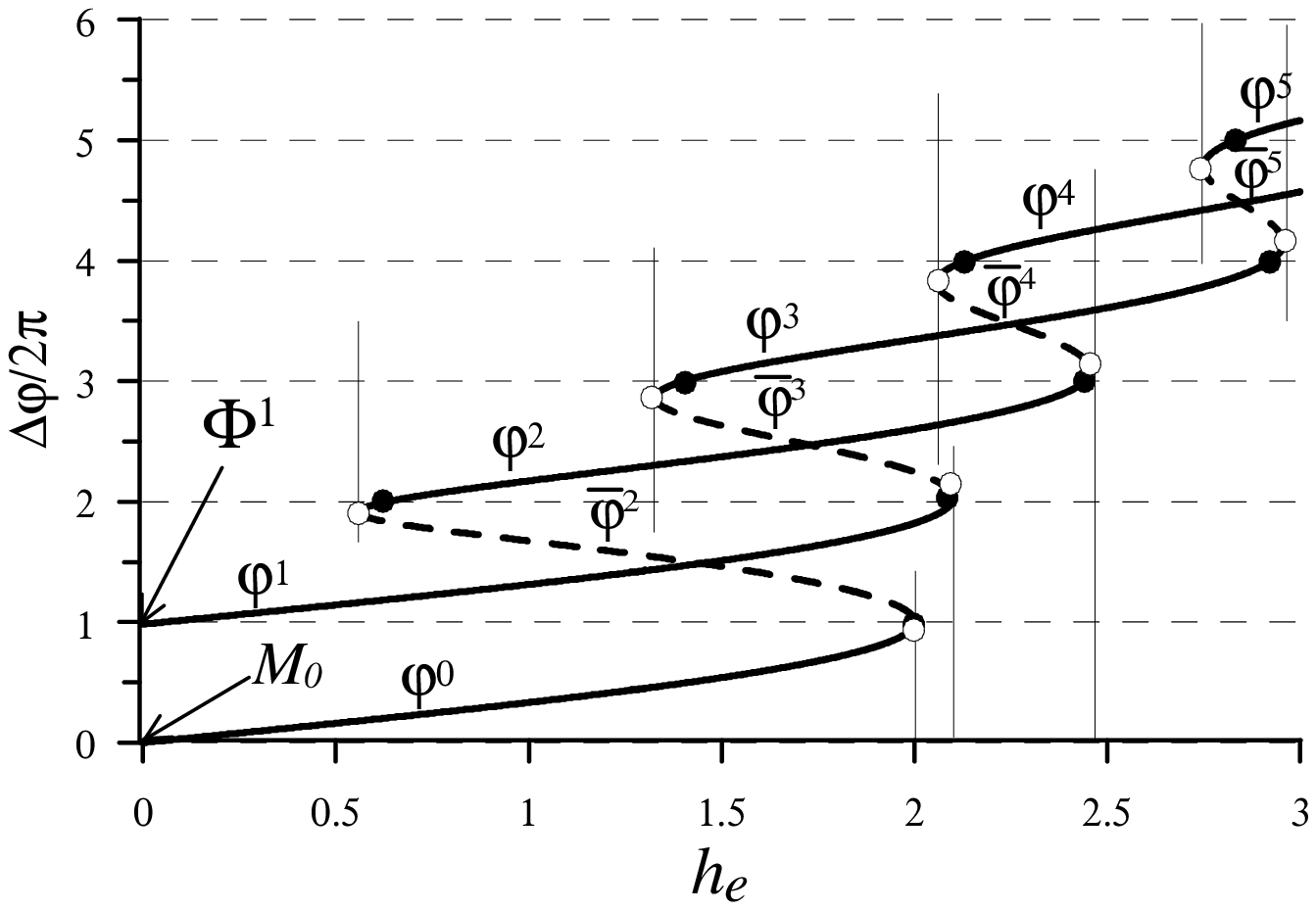}{0.8}{Dependence of the full magnetic flux $\Delta \vp$ on the magnetic field $h_e$ for fluxon distributions associated with $M_0$ and $\Phi^1$. Solid and dashed lines correspond, respectively, stable and unstable states. Light circles indicate the turning points, dark circles indicate the points where solution changes its stability. \label{contin_fig}}
\myfigures{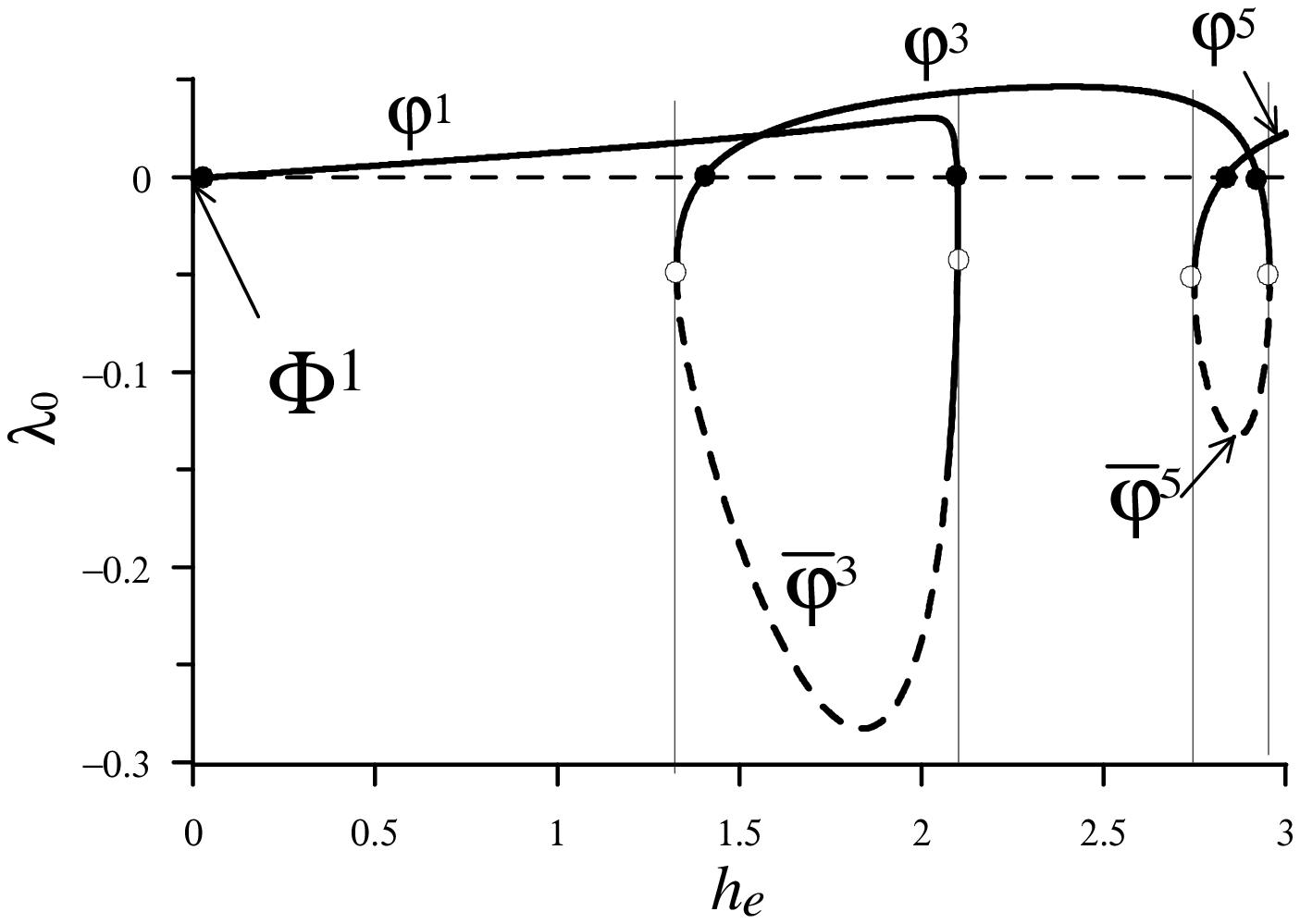}{0.48}{Dependence of the minimal eigenvalue $\lambda_0$  on the magnetic field $h_e$ for the branch associated with $\Phi^1$. Light circles indicate the turning points, dark circles indicate the points of stability change.\label{evfig1}}{0.48}
{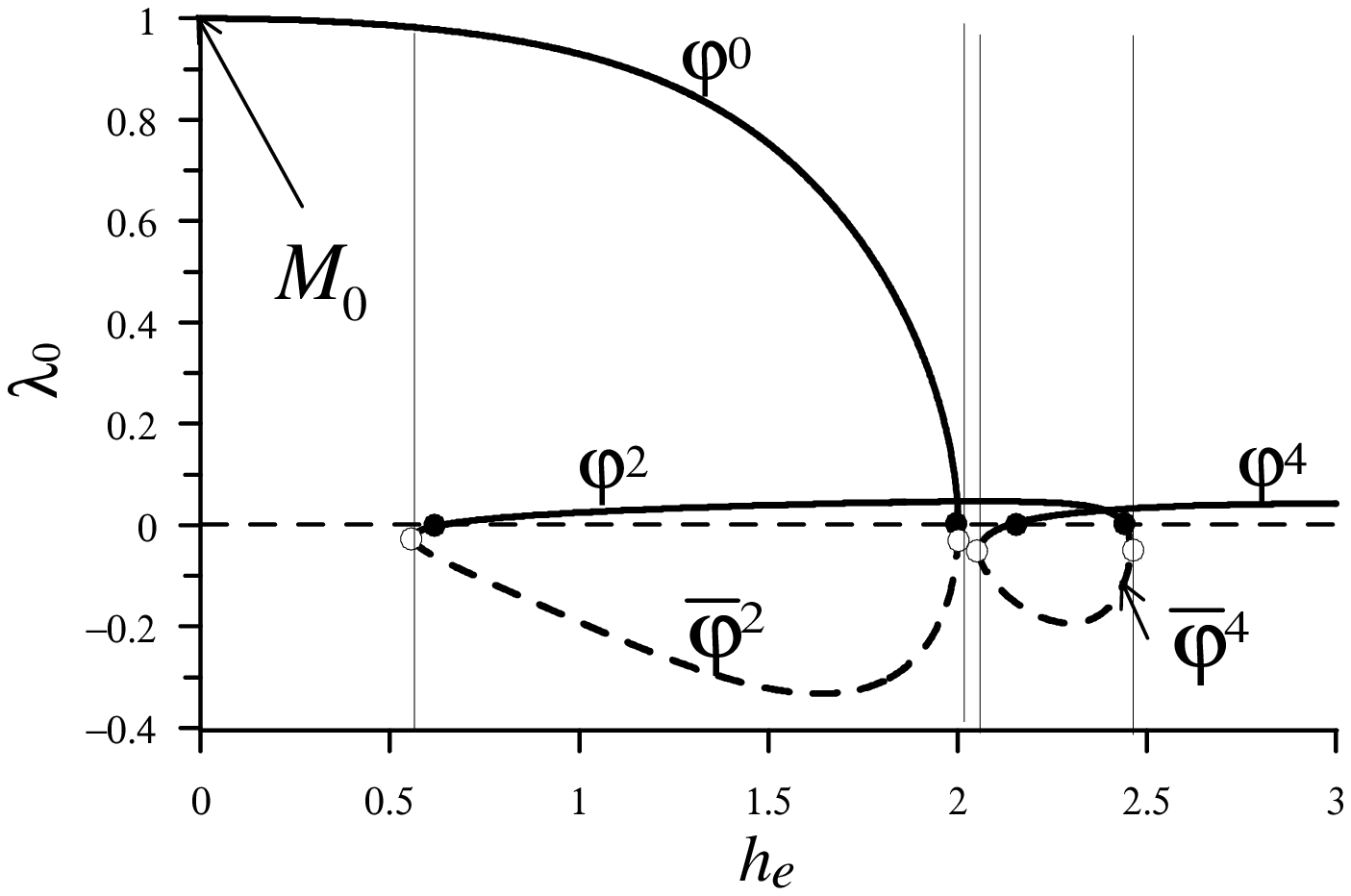}{0.48}{Dependence of the minimal eigenvalue $\lambda_0$ on the magnetic field $h_e$ for the branch associated with $M_0$. Light circles indicate the turning points, dark circles indicate the points of stability change.\label{evfig2}}{0.48}

%
%
 \myfigures{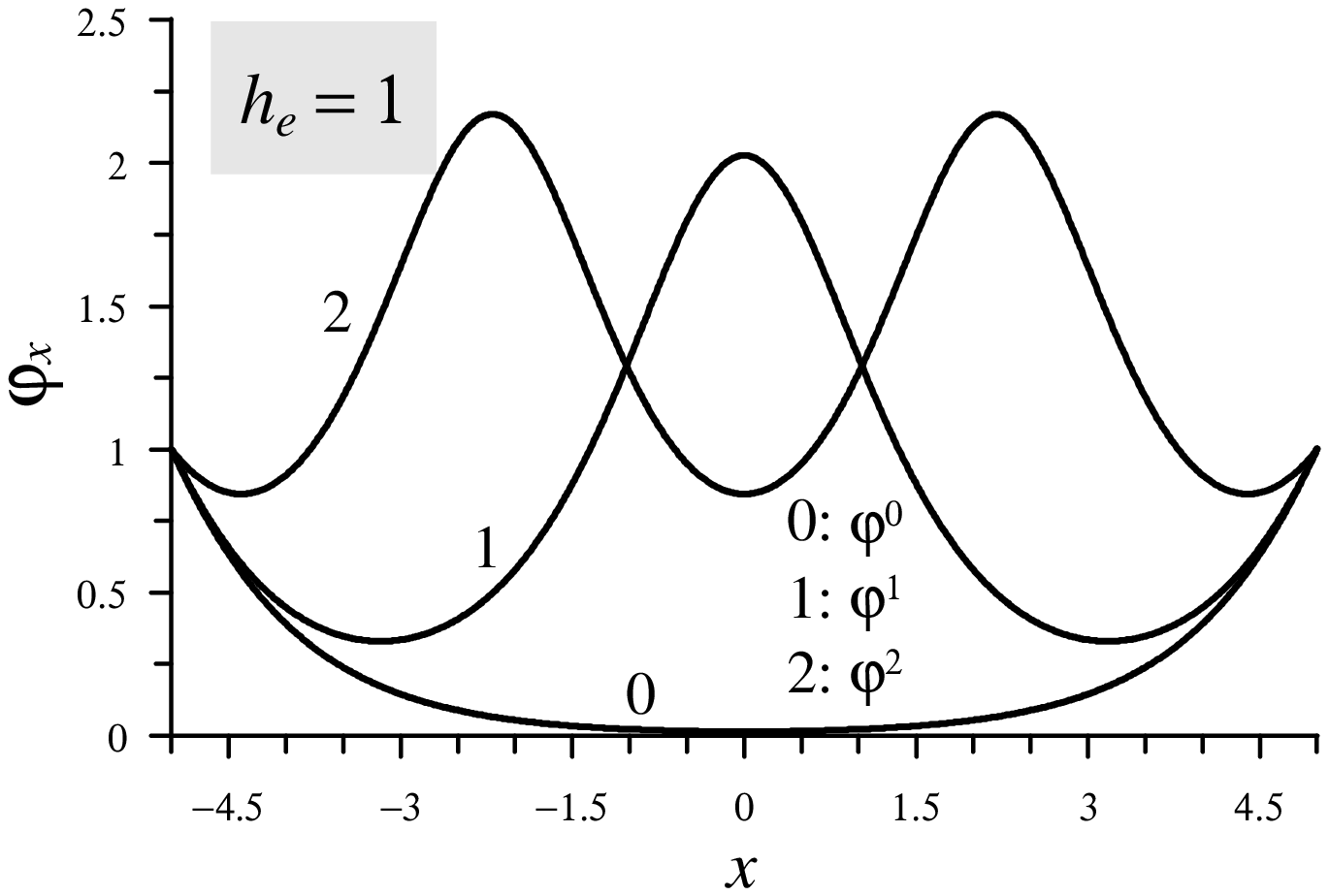}{0.48}{Coexisting stable internal magnetic field distributions $\vp_x$  at magnetic field $h_e = 1$. \label{he1fig2}}{0.48}
{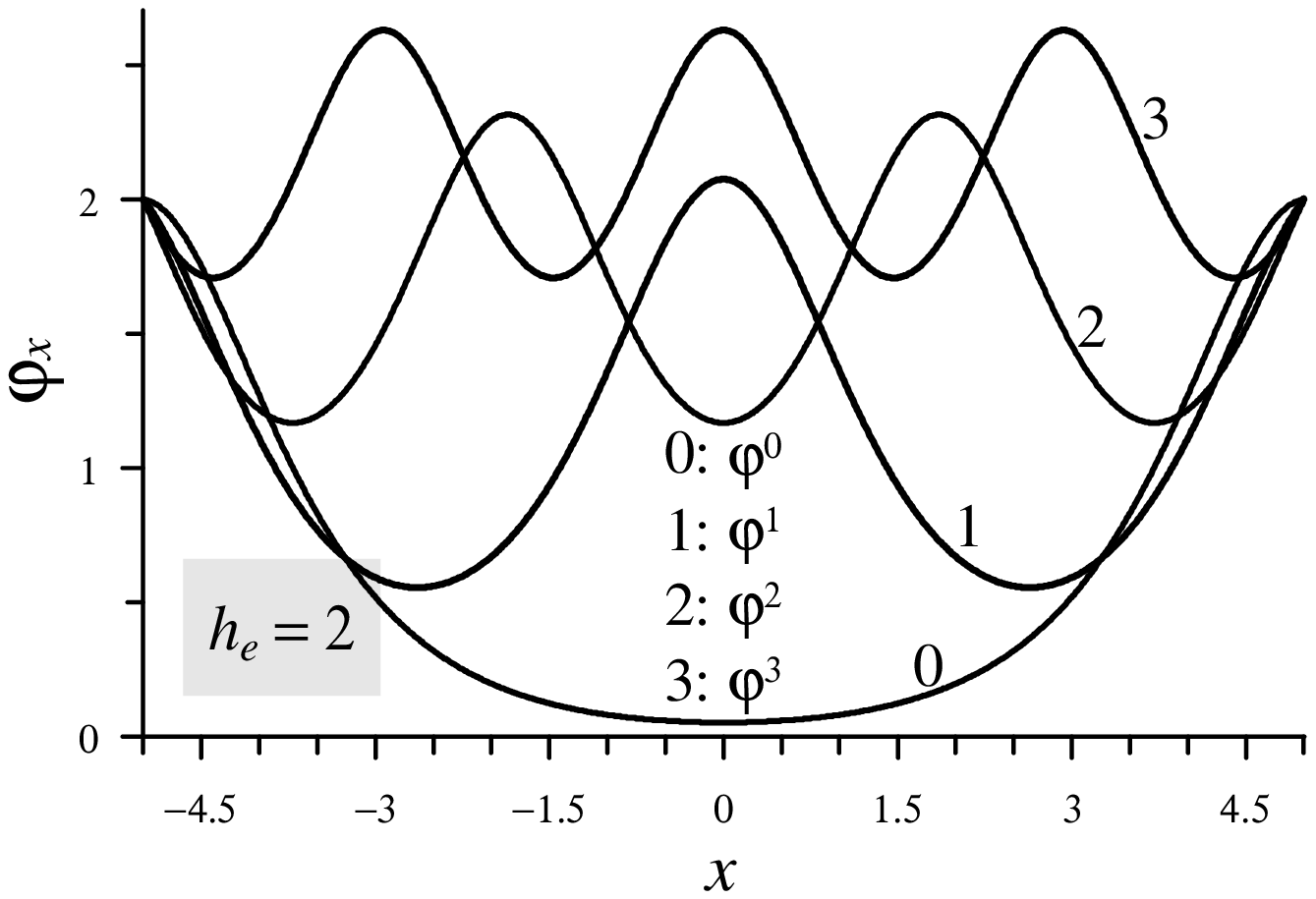}{0.48}{Coexisting stable internal magnetic field distributions $\vp_x$  at magnetic field $h_e = 2$. \label{he2fig2} }{0.48}

   Two {\it basic} distributions are known at $h_e = 0$: the uniform Meissner solution $M_0$ {with $N[M_0] = 0$} and the fluxon solution  $\Phi^1$ {with $N[\Phi^1] = 1$} (\cite{azbsh10}). Minimal eigenvalue $\lambda_0$ of Eq.({\ref{slp}) is negative for $\Phi^1$ and positive for $M_0$.
   As we continue basic state $M_0$ to $h_e>0$ $\lambda_0$ stays positive, i.e. the branch is stable  until  $h_e = 2$. In the $\Phi^1$ case, the minimal eigenvalue $\lambda_0$ crosses zero at the point $h_e=h_1=0.054$, i.e. the branch is unstable for $0\leq h_e\leq h_1$ and stable for $h_1<h_e<2.098$.
   Transformation of the internal magnetic field shape of basic solutions in dependence on $h_e$ is shown on Figs.\ref{solsM0}, \ref{solsF1}.

      The $\Delta\vp(h_e)$ branches associated with the basic solutions $M_0$ and $\Phi^1$ are presented on Fig.\ref{contin_fig}. It is seen, at some  points (light circles in Fig.\ref{contin_fig}) the curves $\Delta\vp(h_e)$ turn back to another, upper,  branches. When the $\Delta\vp(h_e)$ curve turns to the left (``$\supset$''-point) the quantity $N$ is increased to $N+2$. So, the branch started from the basic $M_0$ solution at $h_e = 0$, joins  the fluxons (stable and unstable) with the even $N$ while the another branch (associated with the $\Phi^1$ basic fluxon) connects fluxons (stable and unstable) with the odd $N$.

      The change  of stability occurs at the points (marked by dark circles in Fig.\ref{contin_fig}) where the $\lambda_0(h_e)$ curve crosses zero, see  Figs.\ref{evfig1},\ref{evfig2}. The ``${\supset}$''- and ``$\subset$''-turning points are indicated by the light circles. The ``$\subset$''-turning points connect a pair of unstable solutions with the same number $N$: $\vp^n$ and $\bar\vp^n$. An increasing $N$ to $N+2$ is observed at  the ``${\supset}$''-turning points (light circles).

     Thus, for $0<h_e<h_1$ we have a single stable static distribution (associated with the basic solution $M_0$). For $h_1<h_e<h_{cr}$, $h_{cr}=0.561$ this distribution coexist with another one associated with the basic solution $\Phi^1$. An increasing of magnetic field $h_e$ leads appearing, at $h_{cr}$, (most left light circle in Fig.\ref{contin_fig}) a pair of (unstable) states ($\vp^2$, $\bar\vp^2$). As $h_e$ is growing next, the stabilization of $\bar\vp^2$ occurs (most left dark circle
     in Fig.\ref{contin_fig}), i.e. for $h_e = 1$  we have three stable distributions to be coexisting with  unstable state $\vp^2$, see Fig.\ref{he1fig2}. Further increasing $h_e$ induces a creation, at each ``$\subset$''-point, of additional pair ($\vp^n$, $\bar\vp^n$) with growing $n$, see Fig.\ref{he2fig2}.  At the same time, the pairs  $(\vp^n, \bar{\vp}^{n+2})$ with previous values  $n$  are sequentially disappearing at the ``$\supset$''-points.

\section{Conclusions}
The detailed information on the variation of fluxon structure with external magnetic field in long Josephson
junction is very important for correct interpretation of the experimental results. In this paper  we
investigated stationary fluxon solutions of Eq.(\ref{dinamic_equation_s}),(\ref{postanovka_gr_usl_s}) in
dependence on the external magnetic field $h_e$. Our numerical technique allowed us to establish the
interconnection between the basic solution $M_0$ at $h_e=0$ and the stationary distributions $\vp^n$ with even
numbers $n$ as well as the interconnection between basic state $\Phi^1$ at $h_e=0$ and  $\vp^n$ with odd numbers
$n$. Coexistence of different stable $n$-fluxon distributions  at different values of external magnetic field
$h_e$ is been shown. We consider that predicted transformations of the stable fluxon distributions can be
observed experimentally by investigation of the critical current in dependence of external magnetic field.}

\subsubsection*{Acknowledgments.} The work  was supported by the Program for collaboration of JINR (Dubna) and
Bulgarian scientific centers. EVZ was partially supported by RFBR (Grant No. 09-01-00770). PKhA was partially supported by project NI11-FMI-004.


\end{document}